\documentclass[final,5p,times,twocolumn]{elsarticle} 
\usepackage{graphicx}
\usepackage{amssymb}
\usepackage{xspace}
\usepackage{textcomp}
\usepackage{hyperref}

\begin{document}

\begin{frontmatter}

\title{The MuPix System-on-Chip for the Mu3e Experiment}


\author[PIHeidelbergAddress]{Heiko~Augustin}

\author[MainzAddress]{Niklaus~Berger}

\author[PIHeidelbergAddress]{Sebastian~Dittmeier}

\author[MainzAddress]{Carsten~Grzesik}

\author[PIHeidelbergAddress]{Jan~Hammerich}

\author[MainzAddress]{Qinhua~Huang}

\author[PIHeidelbergAddress]{Lennart~Huth}

\author[PIHeidelbergAddress]{Moritz~Kiehn}

\author[MainzAddress]{Alexandr~Kozlinskiy}

\author[PIHeidelbergAddress]{Frank~Meier~Aeschbacher}

\author[KITAddress]{Ivan~Peri\'c}

\author[PIHeidelbergAddress]{Ann-Kathrin~Perrevoort}

\author[PIHeidelbergAddress]{Andr\'{e}~Sch\"{o}ning}

\author[PIHeidelbergAddress]{Shruti~Shrestha\fnref{myfootnote}}

\author[MainzAddress]{Dorothea~vom~Bruch}

\author[MainzAddress]{Frederik~Wauters}

\author[PIHeidelbergAddress]{Dirk~Wiedner\corref{mycorrespondingauthor}}
\cortext[mycorrespondingauthor]{Corresponding author}
\ead{dwiedner@cern.ch}

\address[PIHeidelbergAddress]{Physikalisches Institut der Universit\"{a}t Heidelberg, INF 226, 69120 Heidelberg, Germany}
\address[MainzAddress]{Institut f\"{u}r Kernphysik, Johann-Joachim-Becherweg 45, Johannes Gutenberg-Universit\"{a}t Mainz, 55128 Mainz, Germany}
\fntext[myfootnote]{Now at Department of Physics, Kansas State University, 116, Cardwell Hall, Manhattan, KS, 66506, USA}
\address[KITAddress]{Institut f\"{u}r Prozessdatenverarbeitung und Elektronik, KIT,\\ Hermann-von-Helmholtz-Platz 1, 76344 Eggenstein-Leopoldshafen, Germany}

\begin{abstract}

Mu3e is a novel experiment searching for charged lepton flavor violation in the rare decay $\mu^+ \rightarrow e^+e^-e^+$. Decay vertex position, decay time and particle momenta have to be precisely measured in order to reject both accidental and physics background. A silicon pixel tracker based on 50\,\textmu m thin high voltage monolithic active pixel sensors (HV-MAPS) in a 1\,T solenoidal magnetic field provides precise vertex and momentum information. The MuPix chip combines pixel sensor cells with integrated analog electronics and a periphery with a complete digital readout. The MuPix7 is the first HV-MAPS prototype implementing all functionalities of the final sensor including a readout state machine and high speed serialization with 1.25\,Gbit/s data output, allowing for a streaming readout in parallel
to the data taking. The observed efficiency of the MuPix7 chip including the full readout system is $\geq99\%$ in a high rate test beam.
\end{abstract}


\end{frontmatter}


\section{The Mu3e Experiment}

The Mu3e experiment~\cite{RP} will be carried out at the Paul Scherrer Institute (PSI) in Switzerland aiming to find or exclude the decay $\mu^+ \rightarrow e^+e^-e^+$\xspace at a sensitivity level in branching ratio of $10^{-16}$, improving the current limit of $10^{-12}$ established by SINDRUM in 1988~\cite{Bellgardt:1987du} by four orders of magnitude. This requires operating the Mu3e experiment at very high muon stopping rates of $\mathcal{O}$($10^{9}$\,Hz) for several years while keeping the background below the $10^{-16}$ level. In particular, background from radiative muon decays with internal conversion $\mu^+ \rightarrow e^+e^-e^+\nu\bar{\nu}$\, needs to be suppressed by an excellent momentum resolution well below 0.5 MeV/c. Accidental background can be suppressed by a combination of precise time resolution of $\mathcal{O}(100\,$ps$)$, vertex resolution of $\mathcal{O}$(200\,\textmu m) and a very good momentum resolution. Since electrons and positrons from the muon decay at rest have a maximum momentum of 53\,MeV/c, the momentum and vertex resolutions are limited by multiple Coulomb scattering. This requires the reduction of the detector material to 0.l\,\% of a radiation length X$_{0}$ per detector layer.

\begin{figure}[tbp] 
\centering
\includegraphics[width=0.45\textwidth]{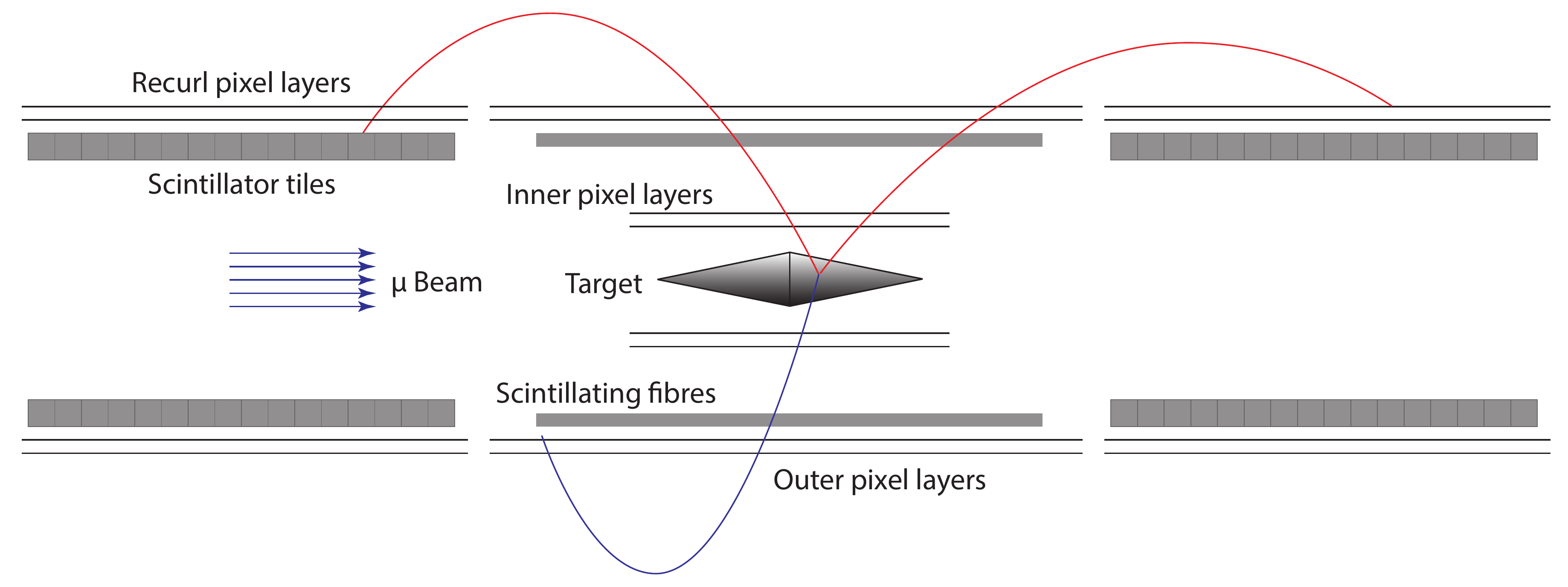}
\caption[Mu3e detector with signal event]
	{Mu3e detector with signal event. 
	}
\label{fig:Mu3eSchematic}
\end{figure}

The Mu3e detector is composed of an extremely lightweight silicon pixel detector~\cite{Berger:2013raa} surrounding a double cone target in a magnetic field of 1\,T in combination with a scintillating fiber and a scintillating tile detector~\cite{Damyanova2014, Eckert2015} for precise timing, see Fig.~\ref{fig:Mu3eSchematic}. The silicon pixel detector is based on high voltage monolithic active pixel sensors (HV-MAPS) thinned to 50\,\textmu m, readout via aluminium flex prints and mounted on Kapton\textsuperscript{\textregistered} frames.  

\section{High Voltage Monolithic Active Pixel Sensors}

\begin{figure}[tbp] 
\centering
\includegraphics[width=0.45\textwidth]{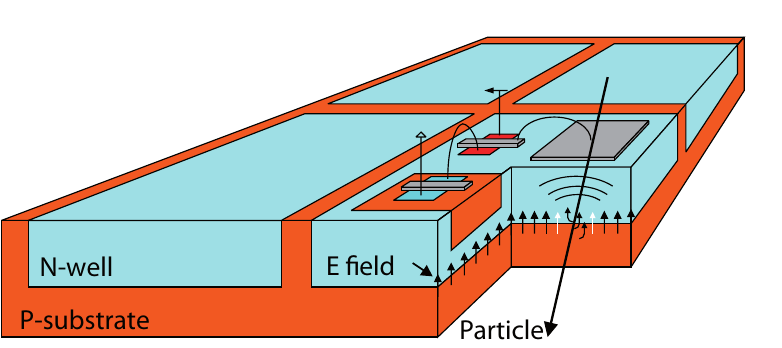}
\caption[HV-CMOS Pixel Scheme]
	{HV-CMOS pixel detector scheme~\cite{Peric:2007zz}.
	}
\label{fig:HV-CMOS-Sketch}
\end{figure}

High voltage monolithic active pixel sensors~\cite{Peric:2007zz, Peric:2015ska} are pixelated detectors based on the commercially available HV-CMOS technology. Reverse biasing the deep N-well in the P-substrate with -60\,V to -85\,V leads to a depletion zone in the order of 15\,\textmu m thickness as reported in~\cite{Peric:2015ska}. Ionizing radiation creates electron-hole pairs in the depletion zone. The electrons are collected via drift within 1\,ns, see Fig.~\ref{fig:HV-CMOS-Sketch}. Because of the very thin active detection layer, thinning to 50\,\textmu m is possible. In addition, CMOS analog and digital electronics can be implemented in the N-well so no extra readout chip is required.

\section{The MuPix System on Chip}

For the MuPix chip to be used in the Mu3e experiment, a pixel size of 80 $\times$ 80\,\textmu m$^2$, an overall dimension of 2 $\times$ 2\,cm$^2$ and a thickness of only 50\,\textmu m is required.

\begin{figure}[tbp] 
\centering
\includegraphics[width=0.50\textwidth]{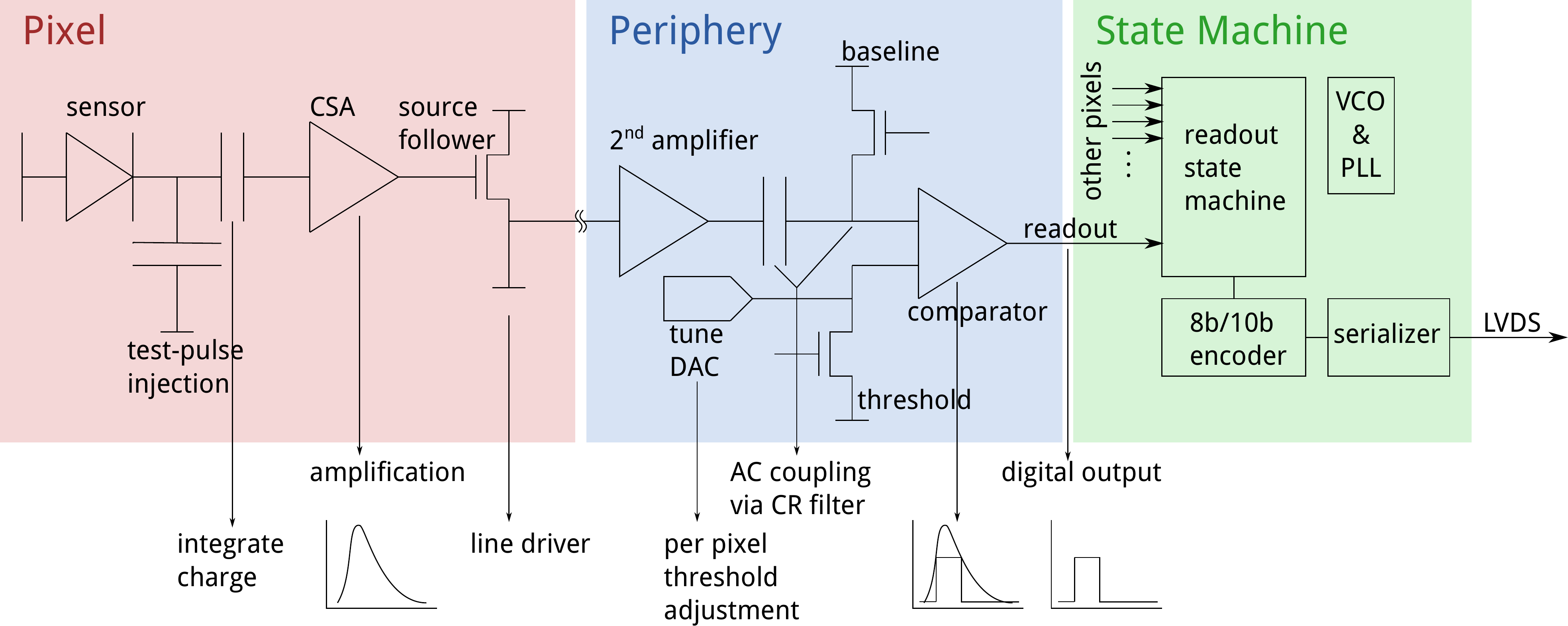}
\caption[Block diagram of MuPix signal path]
	{Block diagram of the signal path in the MuPix7 chip. 
	}
\label{fig:PixelandPeriphery}
\end{figure}

\begin{figure}[tbp] 
\centering
\includegraphics[width=0.45\textwidth]{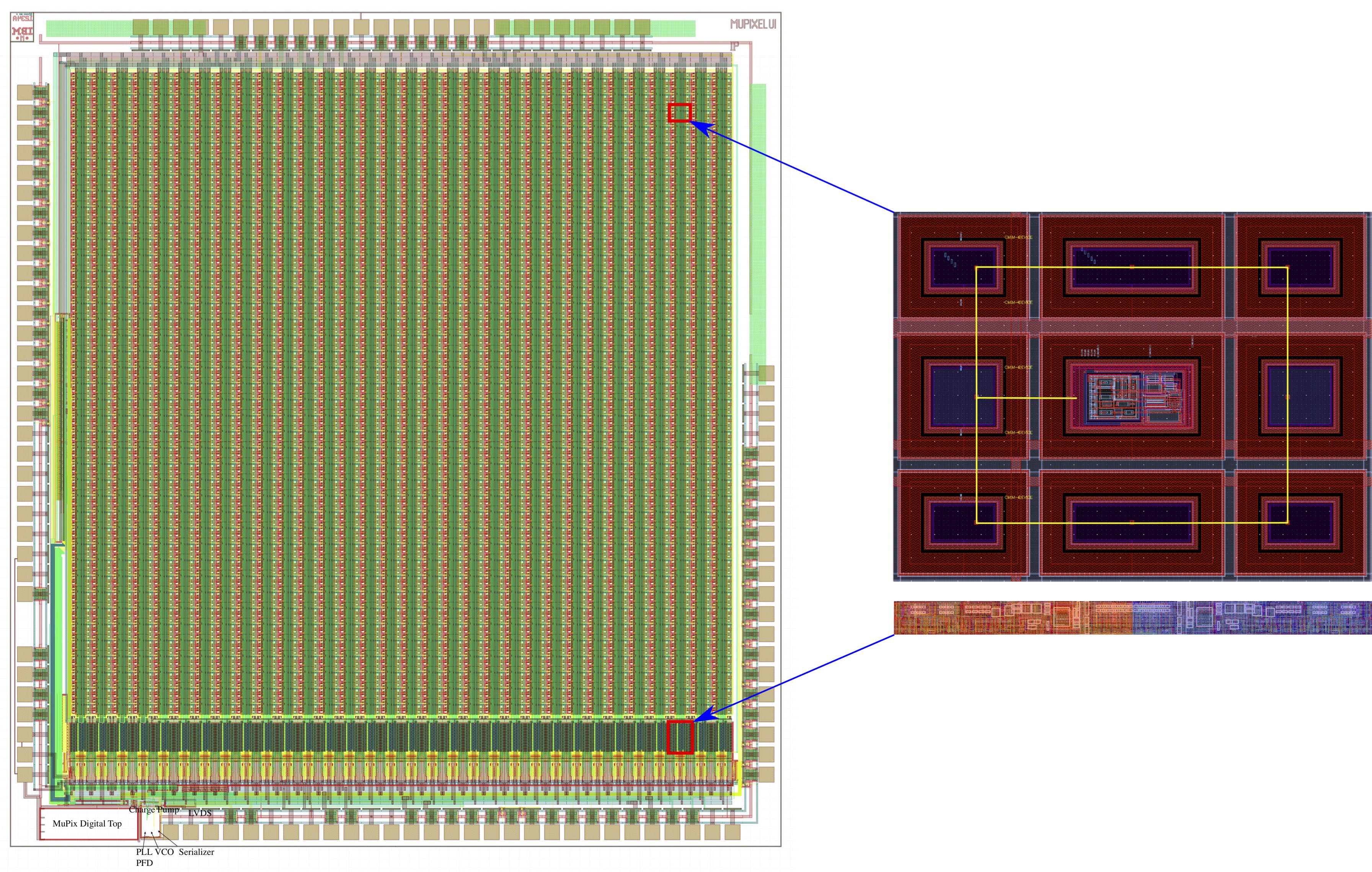}
\caption[MuPix7 layout picture]
	{MuPix7 layout and zoom-in of one pixel and the corresponding electronics in the periphery. The pixel cell is subdivided in nine diodes with the amplifier electronics implemented in the central one. Note that periphery electronics serving two pixels is shown.
	}
\label{fig:MuPix7}
\end{figure}

Several smaller prototypes implemented in the AMS 180\,nm HV-CMOS process have been extensively characterized~\cite{Shrestha:2014oxa, Augustin:2015mqa, Augustin2014, Kiehn2015}. A line driver in the pixel cell sends the analog pulse over a point-to-point connection to a corresponding digital pixel in the periphery. The signal is discriminated, a hit flag is registered and an eight-bit time-stamp is latched. Additionally, each digital pixel has a four-bit tune digital to analog converter (DAC) which allows to adjust the baseline for a common threshold in order to correct for pixel to pixel variations, see Fig.~\ref{fig:PixelandPeriphery}. The analog part with the preamplifier is implemented in the deep N-well above the sensitive pixel area, while the digital part is located in a small inactive region next to the pixel matrix in order to avoid digital crosstalk, see Fig.~\ref{fig:MuPix7}.

The MuPix7 prototype is the first in its series which integrates the complete readout circuitry, including a readout state machine, fast clock circuitry and fast serial output. 
At the beginning of each readout cycle, the hit flags of all pixels are copied to a second register. These registers drive a priority logic selecting the first hit in each column, which is then copied to a buffer in the column periphery. At the same time, the pixel hit flag is cleared and set ready to detect the next hit. A second priority logic identifies the first column with a hit, which is then sent to the fast output link. This is repeated until all column buffers are empty and the next hit in each column can be copied to the periphery. If all hits are read or an adjustable number of hits is surpassed, the cycle starts from the beginning. This readout is controlled by an on chip state machine running at 62.5\,MHz and runs in parallel to the data taking. The only dead time incurred is in the hit pixels waiting for the copying to the column periphery. At low occupancy, this is comparable to the shaping time of around 1\,\textmu s.
The output data consists of the row, column address and Gray-encoded time-stamp of hits interspersed with control words and synchronization counters. The data is 8b/10b encoded, serialized and sent off-chip via a 1.25\,Gbit/s low voltage differential signaling (LVDS) link. The clocks required for the time-stamp generation, the readout state machine, the serializer and the fast link are all generated on-chip. To this end a voltage controlled oscillator together with a phase locked loop connected to an external 125\,MHz clock have been implemented. The MuPix7 prototype has 32 by 40 pixels of 103 $\times$ 80\,\textmu m$^2$ size and was characterized in numerous laboratory tests and test beam campaigns.

The MuPix7 is directly glued and bonded to a dedicated test board, the MuPix7 printed circuit board (PCB), which delivers ripple-free and stabilized low voltage, high voltage, threshold and test pulses to the chip. A commercially available state-of-the-art FPGA (ALTERA Stratix IV) on a PCIe card is used for generating the master clock and handling the communications to and from the MuPix7 chip. LVDS is used for all signals with speeds of up to 1.25\,Gbit/s. Four planes of MuPix7 PCBs were combined in a MuPix beam telescope~\cite{Huth2014}, with first, third and fourth MuPix7 providing the track reference and the second MuPix7 as device under test (DUT).   

\section{MuPix7 Performance}

\begin{figure}[tbp] 
\centering
\includegraphics[width=0.45\textwidth]{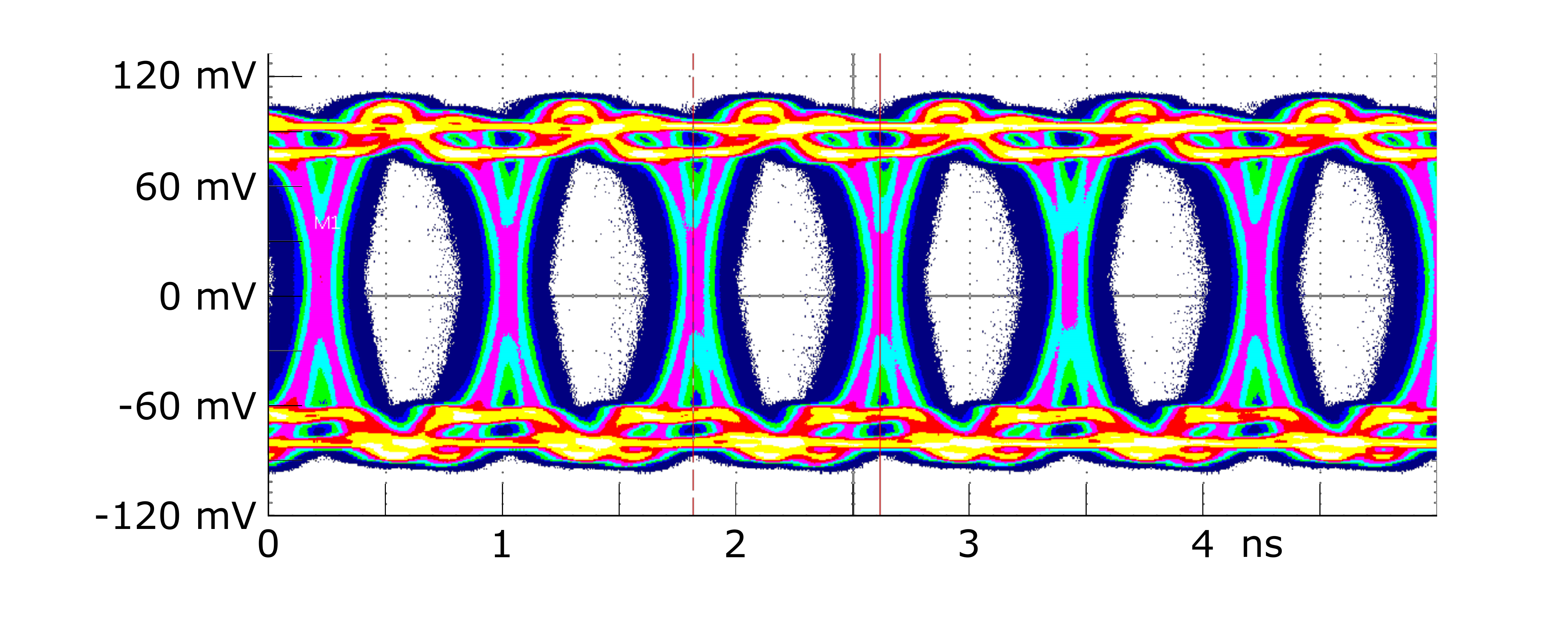}
\caption[MuPix7 eye diagram of fast serial data output]
	{Eye diagram of MuPix7 fast serial data output, shown as signal amplitude in mV versus time in ns.  
	}
\label{fig:MuPix7-eye-diagram}
\end{figure}

The MuPix7 chip is the first HV-CMOS chip with an internal readout and a fast 1.25\,Gbit/s LVDS serial data output. The high speed serial output data shows very good signal quality, see Fig.~\ref{fig:MuPix7-eye-diagram}. The eye opening has 131\,mV height with 5.4\,mV RMS noise and 525\,ps width with 45\,ps jitter when probed with high bandwidth Sub-Miniature-A (SMA) cables close to the chip outputs.

High hit rate tests up to 380\,kHz per sensor have been performed at PSI with a 250\,MeV/c mixed positron, muon and pion beam. At the CERN SPS, a 180\,GeV/c pion beam with an instantaneous hit rate of above 500\,kHz per sensor plane was used for testing. Further studies have been carried out with a 1\,GeV/c electron beam of MAMI at the Johannes Gutenberg University Mainz and a 2 to 6\,GeV/c positron beam at DESY - each with a hit rate of around 1\,kHz. 

The rate capability tests were performed at the PSI with hit rates between 0.15\,kHz and 380\,kHz. At 380\,kHz the chip was using only about $4\%$ of its available bandwidth. The theoretical upper limit for the MuPix7 chip is a hit rate of 30\,MHz corresponding to a flux of 284\,MHz/cm$^2$. For the expected rate of $2\cdot10^{9}$ muon stops per second at the Mu3e experiment the charged particle flux will be up to 40\,MHz/cm$^2$. A full sized chip will have up to four fast serial output links.

\begin{figure}[tbp] 
\centering
\includegraphics[width=0.45\textwidth]{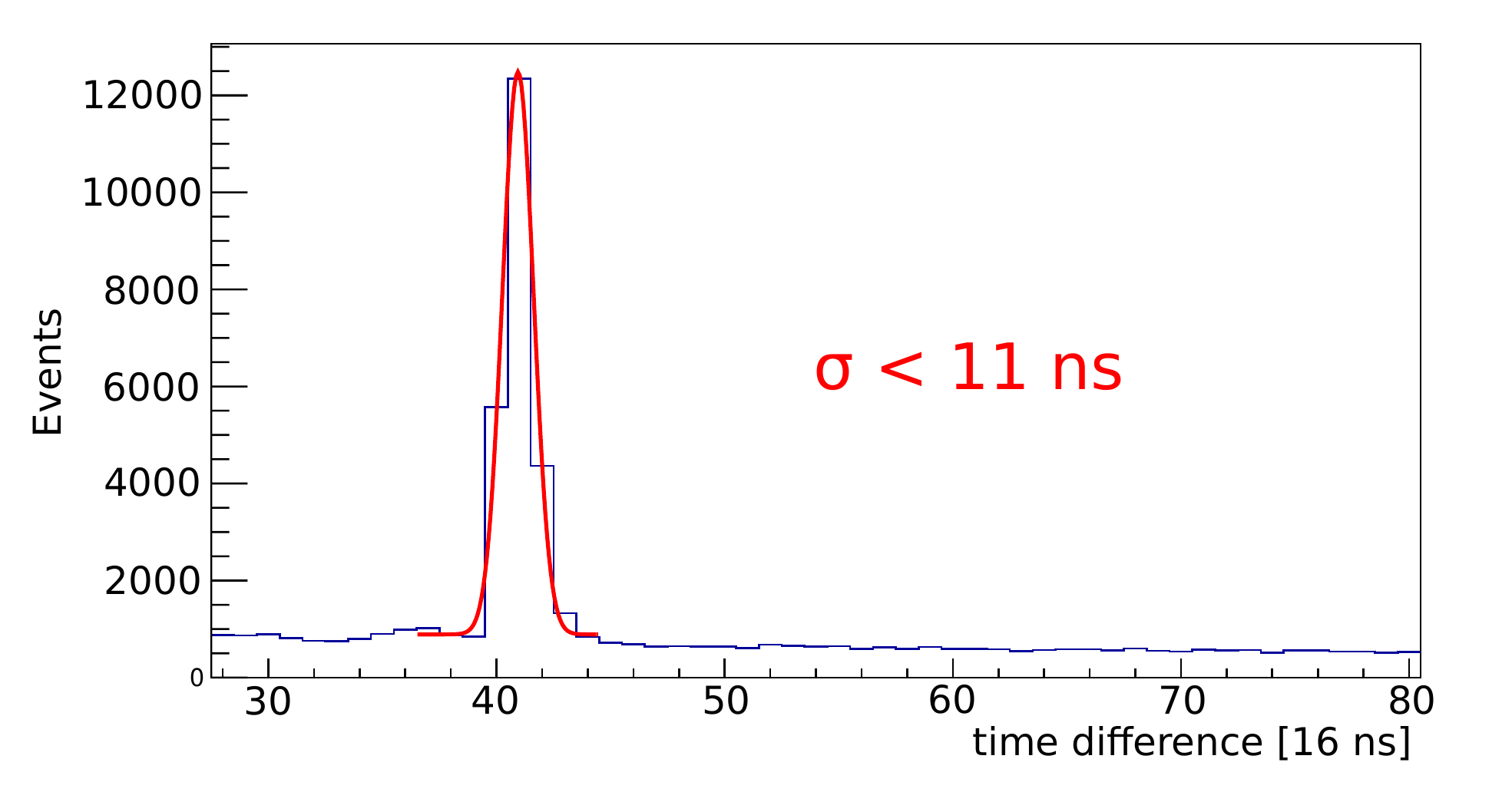}
\caption[MuPix7 time resolution]
	{Time resolution of the MuPix7 chip.
	}
\label{fig:MuPix7-time-resolution}
\end{figure}

The time resolution of the MuPix7 has been determined by sampling with a 62.5\,MHz Gray counter, see Fig.~\ref{fig:MuPix7-time-resolution}. Plastic scintillators were used as time reference. A Gaussian fit of the relative time between MuPix7 hit time and scintillator signals yields $\sigma=$11\,ns, which gives an upper limit of the time resolution. 

\begin{figure}[tbp] 
\centering
\includegraphics[width=0.45\textwidth]{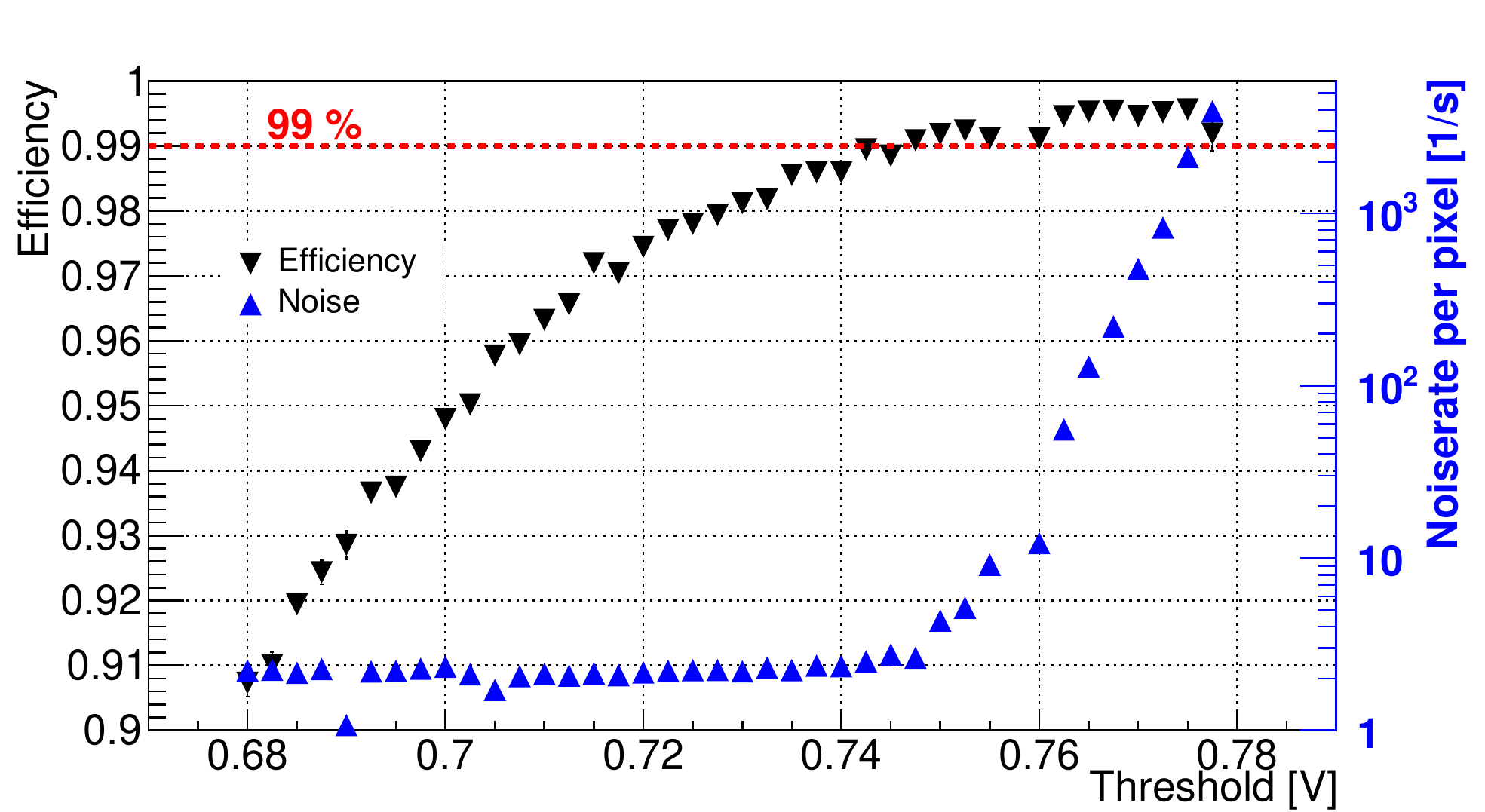}
\caption[MuPix7 hit efficiencies versus threshold]
	{Hit efficiencies and noise versus threshold for the MuPix7.
	}
\label{fig:MuPix7-hit-efficiencies}
\end{figure}

For the measurement of the particle detection efficiency, the noise of each pixel has been equilibrated to 1\,Hz for one global threshold with the help of the tune DACs and a threshold scan has been performed. In the plateau, hit efficiencies of around $99.5\%$ have been determined for the 250\,MeV mixed positron, muon and pion beam at PSI, see Fig.~\ref{fig:MuPix7-hit-efficiencies}. The noise in the efficiency plateau region is in the range of 2 to 12\,Hz per pixel, with a power consumption of around 300\,mW/cm$^2$, which is within the Mu3e pixel detector cooling budget of 400\,mW/cm$^2$. 

\begin{figure}[tbp] 
\centering
\includegraphics[width=0.45\textwidth]{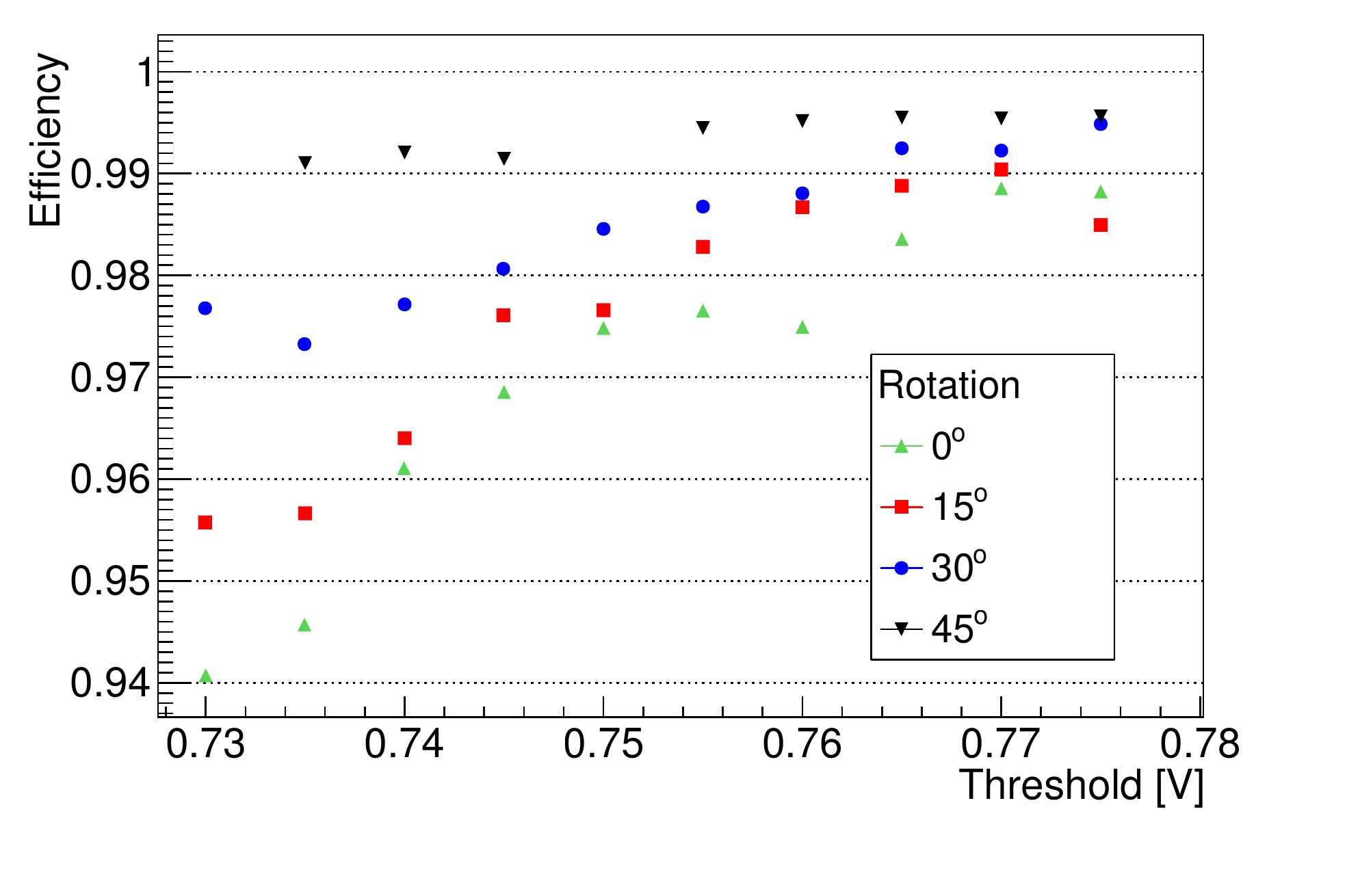}
\caption[MuPix7 hit efficiencies rotated sensor]
	{Hit efficiencies versus threshold for the MuPix7 chip at angles of 0$^\circ$, 15$^\circ$, 30$^\circ$ and 45$^\circ$ with respect to the beam axis. Statistical errors are smaller than the markers.
	}
\label{fig:MuPix7-hit-efficiencies-rotated-sensor}
\end{figure}

In order to study the dependency of the detection efficiency on the thickness of the depletion zone, threshold scans at four different angles with respect to the positron beam were conducted, using a 4\,GeV/c positron beam at DESY, see Fig.~\ref{fig:MuPix7-hit-efficiencies-rotated-sensor}. The measurement taken at an angle of 45$^\circ$ shows a much broader efficiency plateau as compared to the 0$^\circ$ measurement. 45$^\circ$ corresponds to a $\sqrt{2}$ times thicker effective depletion zone. A planned change of the substrate resistivity from 20\,$\Omega$cm to 80\,$\Omega$cm should lead to an increase in number of signal electrons by a factor of two and thus an even broader efficiency plateau.

\begin{figure}[tbp] 
\centering
\includegraphics[width=0.45\textwidth]{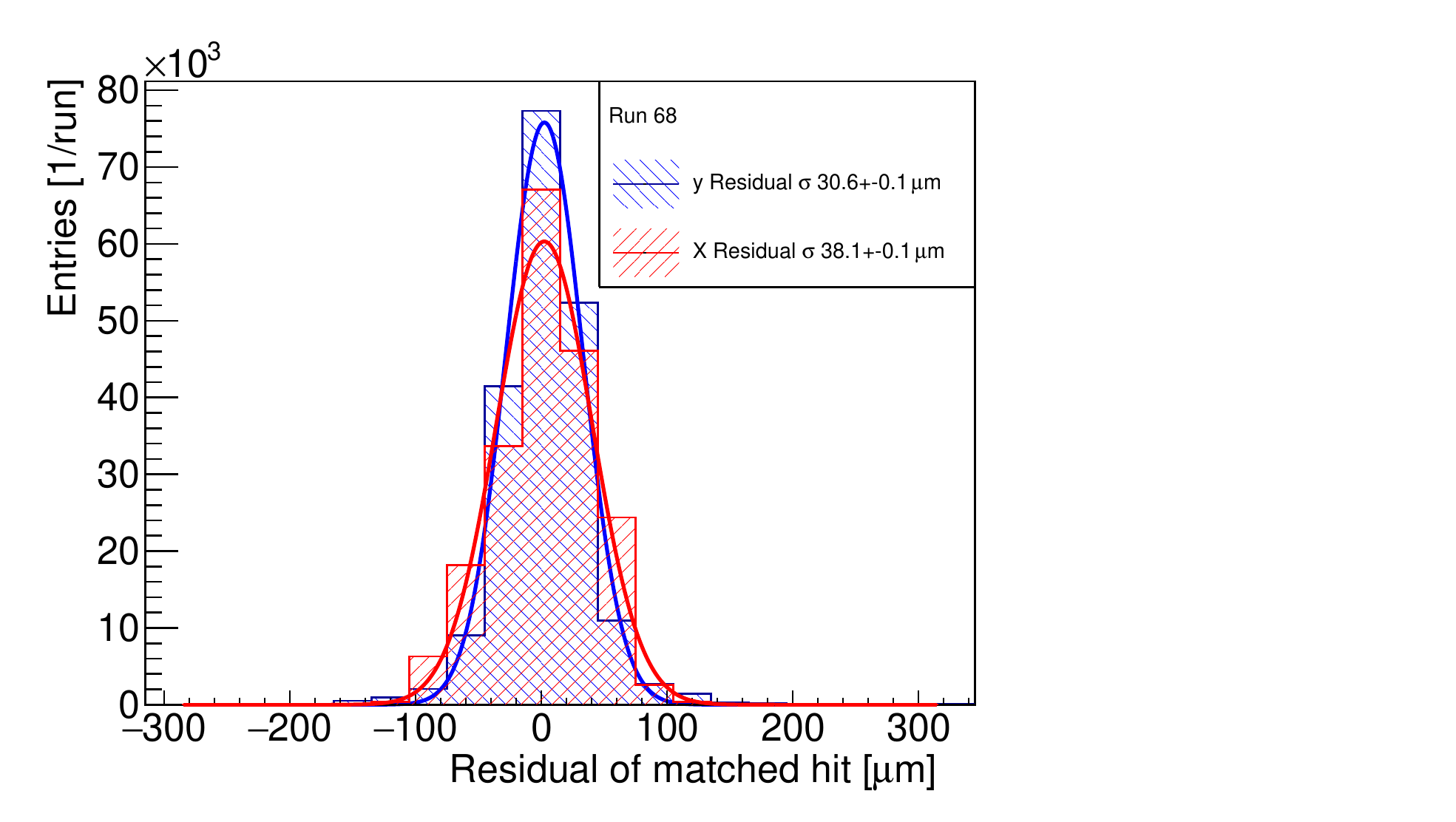}
\caption[MuPix7 spatial resolution]
	{Spatial resolution of the MuPix7 chip.
	}
\label{fig:MuPix7-spatial-resolution}
\end{figure}

The spatial resolution of the MuPix7 chip has been measured using a 4\,GeV/c positron beam at DESY and a MuPix beam telescope, see Fig.~\ref{fig:MuPix7-spatial-resolution}. The unbiased hit residuals are $\sigma_x$=38.1\,\textmu m and $\sigma_y$=30.6\,\textmu m, which is consistent with the intrinsic single cell resolution (pitch/$\sqrt{12}$) folded with uncertainties from multiple Coulomb scattering and tracking. 
All studies above were carried out with MuPix7 samples thinned to 50\,\textmu m or 62\,\textmu m. No influence of the thickness on the performance was observed. 

\section{Conclusions}

The current prototype MuPix7 is a 50\,\textmu m thin full system on chip combining pixel sensor, analog electronics and full digital readout. All pixels and the integrated readout circuitry are always running in parallel, so no dead-time on top of the signal shaping time of around 1\,\textmu s is induced. The newly developed 1.25\,Gbit/s fast serial data readout has an excellent performance with large eye openings and has been reliably operated in multiple test beam campaigns. The time resolution of the MuPix7 time stamps is better than $\sigma=$11\,ns. The spatial resolution of $\leq$38.1\,\textmu m is dominated by the pixel cell size contribution. The full system hit efficiency is $\geq99\%$.  

The MuPix7 chip performs very well, meeting all requirements that can be investigated on a small scale prototype. Upscaling to a 13 $\times$ 23\,mm$^2$ pixel chip is the next step, together with an investigation of higher resistivity substrates.

The MuPix7 is a chip demonstrating the unique capabilities of HV-MAPS technology. High rates and lowest amount of traversed material make them the ideal candidate for future applications, enabling precision experiments like Mu3e. The MuPix chip has also been chosen as detector for the PANDA Lumi detector~\cite{Weber2016} and the P2 experiment~\cite{Berger:2015aaa}. Their moderate price allows for relatively large instrumented areas, making them a candidate for track-trigger applications at large radii in LHC-type experiments. Irradiation studies are ongoing to validate the HV-MAPS technology for LHC.

\section*{Acknowledgments}
N.~Berger, Q.~Huang, A.~Kozlinskiy, S.~Shrestha, D.~vom~Bruch and F.~Wauters thank the \textit{Deutsche Forschungsgemeinschaft} for supporting them and the Mu3e project through an Emmy Noether grant. 
S.~Dittmeier, L.~Huth and M.~Kiehn acknowledge support by the IMPRS-PTFS. 
A.-K.~Perrevoort acknowledges support by the Particle Physics beyond the Standard Model research training group [GRK 1940].
H.~Augustin acknowledges support by the HighRR research training group [GRK 2058]. 
N.~Berger and A.~Kozlinskiy thank  the PRISMA Cluster of Excellence for support. 

The measurements leading to these results have been performed at the Test Beam Facility at DESY Hamburg (Germany), a member of the Helmholtz Association (HGF).
We thank the Institut f\"{u}r Kernphysik at the Johannes Gutenberg University Mainz for giving us the opportunity to take data at the MAMI beam. 
We owe our SPS test beam time to the SPS team and our LHCb colleagues, especially Heinrich Schindler, Kazu Akiba and Martin van Beuzekom.
We would like to thank PSI for valuable test beam time.


\bibliography{mybibfile}

\begin{thebibliography}{10}
\expandafter\ifx\csname url\endcsname\relax
  \def\url#1{\texttt{#1}}\fi
\expandafter\ifx\csname urlprefix\endcsname\relax\def\urlprefix{URL }\fi
\expandafter\ifx\csname href\endcsname\relax
  \def\href#1#2{#2} \def\path#1{#1}\fi

\bibitem{RP}
A.~{Blondel}, et~al., {Research Proposal for an Experiment to Search for the
  Decay $\mu \rightarrow eee$ }\href {http://arxiv.org/abs/1301.6113}
  {\path{arXiv:1301.6113}}.

\bibitem{Bellgardt:1987du}
U.~Bellgardt, et~al., {Search for the Decay $\mu^+ \rightarrow e^+ e^+ e^-$},
  Nucl.Phys. B299 (1988) 1.
\newblock \href {http://dx.doi.org/10.1016/0550-3213(88)90462-2}
  {\path{doi:10.1016/0550-3213(88)90462-2}}.

\bibitem{Berger:2013raa}
N.~Berger, et~al., {A Tracker for the Mu3e Experiment based on High-Voltage
  Monolithic Active Pixel Sensors}, Nucl. Instr. Meth. A 732 (2013) 61--65.
\newblock \href {http://arxiv.org/abs/1309.7896} {\path{arXiv:1309.7896}},
  \href {http://dx.doi.org/10.1016/j.nima.2013.05.035}
  {\path{doi:10.1016/j.nima.2013.05.035}}.

\bibitem{Damyanova2014}
A.~Damyanova, {Development of a Scintillating Fibre Tracker/Time-of-Flight
  Detector with SiPM Readout for the Mu3e Experiment at PSI}, Master thesis,
  Universit\'{e} de Gen\`{e}ve (2013).

\bibitem{Eckert2015}
P.~Eckert, {The Mu3e Tile Detector}, Phd thesis, Kirchhoff-Institute for
  Physics Heidelberg University (2015).

\bibitem{Peric:2007zz}
I.~Peri\'c, {A novel monolithic pixelated particle detector implemented in
  high-voltage CMOS technology}, Nucl.Instrum.Meth. A582 (2007) 876.
\newblock \href {http://dx.doi.org/10.1016/j.nima.2007.07.115}
  {\path{doi:10.1016/j.nima.2007.07.115}}.

\bibitem{Peric:2015ska}
I.~Peri\'c, et~al., {Overview of HVCMOS pixel sensors}, JINST 10~(05) (2015)
  C05021.
\newblock \href {http://dx.doi.org/10.1088/1748-0221/10/05/C05021}
  {\path{doi:10.1088/1748-0221/10/05/C05021}}.

\bibitem{Shrestha:2014oxa}
S.~Shrestha, {The High-Voltage Monolithic Active Pixel Sensor for the Mu3e
  Experiment}, PoS TIPP2014 (2014) 047.

\bibitem{Augustin:2015mqa}
H.~Augustin, N.~Berger, S.~Bravar, S.~Corrodi, A.~Damyanova, et~al., {The MuPix
  high voltage monolithic active pixel sensor for the Mu3e experiment}, JINST
  10~(03) (2015) C03044.
\newblock \href {http://dx.doi.org/10.1088/1748-0221/10/03/C03044}
  {\path{doi:10.1088/1748-0221/10/03/C03044}}.

\bibitem{Augustin2014}
H.~Augustin, {Characterization of a novel HV-MAPS Sensor with two Amplification
  Stages and first examination of thinned MuPix Sensors}, Master thesis,
  Physics Institute Heidelberg University (2014).

\bibitem{Kiehn2015}
M.~Kiehn, {Pixel Sensor Evaluation and Track Fitting for the Mu3e Experiment},
  Ph.D. thesis, Physics Institute Heidelberg University (2016).

\bibitem{Huth2014}
L.~Huth, {Development of a Tracking Telescope for Low Momentum Particles and
  High Rates consisting of HV-MAPS}, Master thesis, Physics Institute
  Heidelberg University (2014).

\bibitem{Weber2016}
T.~Weber, {High-Voltage Active Pixel Sensors for the PANDA Luminosity Detector
  and Search for the Decay $Y(4260) \rightarrow \eta_c \eta \pi^+ \pi^-$ at
  BESIII}, Ph.D. thesis, Mainz University (In preparation).

\bibitem{Berger:2015aaa}
N.~Berger, et~al.,
  \href{http://inspirehep.net/record/1404157/files/arXiv:1511.03934.pdf}{{Measuring
  the weak mixing angle with the P2 experiment at MESA}}, in: {10th
  International Workshop on e+e- collisions from Phi to Psi (PHIPSI15) Hefei,
  Anhui, China, September 23-26, 2015}, 2015.
\newblock \href {http://arxiv.org/abs/1511.03934} {\path{arXiv:1511.03934}}.
\newline\urlprefix\url{http://inspirehep.net/record/1404157/files/arXiv:1511.03934.pdf}

\end{thebibliography}

\end{document}